\documentclass[fleqn]{article}

\usepackage[sumlimits]{amsmath}
\usepackage{bbold}                    
\usepackage{amsfonts}
\usepackage{amssymb}
\usepackage{amsopn}
\usepackage{amsthm}                   
\usepackage{mathrsfs}                 
\usepackage{titlesec}
\usepackage{graphicx}
\usepackage{multicol}
\usepackage{cite}

\titleformat{\section}[hang]
 {\large\sffamily\bfseries}
 {\thesection}{.5em}{}
\titleformat{\subsection}[hang]
 {\normalfont\sffamily\bfseries}
 {\thesubsection}{.5em}{}
\titleformat{\subsubsection}[hang]
 {\normalfont\sffamily}
 {\thesubsubsection}{.5em}{}
\titleformat{\paragraph}[runin]
 {\normalfont\sffamily\bfseries}
 {}{.5em}{}

\renewenvironment{figure}%
 {\par\medskip\refstepcounter{figure}\footnotesize\sffamily\selectfont}%
 {\par\medskip}
 {\par\medskip\refstepcounter{cnt_table}\footnotesize\sffamily\selectfont}%
 {\par\medskip}

\def\imu{\mathrm{i}}                  
\DeclareMathOperator{\id}{id}         
\DeclareMathOperator{\tr}{tr}         
\DeclareMathOperator{\Int}{int}       
\newcommand{\e}[1]{\mathrm{e}^{#1}}   
\newcommand{\term}[1]{\textit{#1}}    
\newcommand{\ie}{\textit{i.e.}}       
\newcommand{\eg}{\textit{e.g.}}       
\newcommand{\cf}{\textit{cf.}}        
\newcommand{\etal}{\textit{et al.}}   
\newcommand{\ket}[1]{{|{#1}\rangle}}  

\newcommand{\avr}[1]{\langle{#1}\rangle}
\def\half{\frac{1}{2}}                
\def\thalf{\tfrac{1}{2}}              
\def\idmatrix{\text{\textbb 1}}       
\def\Till{,\!...,}                    
\def\fN{\mathbb{N}}                   
\def\fZ{\mathbb{Z}}                   
\def\fR{\mathbb{R}}                   
\def\intprod
 {{\unitlength 1pt\begin{picture}(9,8)(-7,0)\line(0,1){6}\line(-1,0){6}\end{picture}}}
\newcommand\roplus
 {{\unitlength 1pt\begin{picture}(11,8)(0,0)\linethickness{.3pt}%
  \put(5.0,-.3){\line(0,1){5.4}}\put(2,0){$\ni$}%
 \end{picture}}}
\newcommand\loplus
 {{\unitlength 1pt\begin{picture}(11,8)(0,0)\linethickness{.3pt}%
  \put(5.3,-.3){\line(0,1){5.4}}\put(2,0){$\in$}%
 \end{picture}}}
\def\grG{G}                           
\def\grU{\mathrm{U}}                  
\def\grSO{\mathrm{SO}}                

\newcommand{\figref}[1]{Fig.~\ref{#1}}

\renewcommand{\caption}[1] {\par{\footnotesize\sffamily Figure \thefigure{}: #1}}

\newcommand{\figcaption}[1]{\par\nopagebreak{\textbf{Figure \thefigure{}} #1}}

\newcommand\eprint[1]{{\tt #1}}
\def\refno#1         {\noindent{\footnotesize #1}\par\noindent\rule[5mm]{\textwidth}{0.1mm}\par}
\newenvironment{articlehead}%
 {\begin{quote}\setlength{\parskip}{0pt}\setlength{\parindent}{0pt}}%
 {\end{quote}}
\def\title#1         {{\sffamily\Large\selectfont\textbf{#1}}\\\par\medskip}
\renewcommand{\author}[1]{{\textsc{#1}}}
\def\authore#1#2     {\textsc{#1}\footnote{e-mail: {#2}}\null}

\def\authorae#1#2#3  {\textsc{#1}$^{#2,}$\footnote{e-mail: {#3}}\null}
\def\address#1       {\par{\it #1}}
\renewenvironment{abstract}%
 {\par\medskip\small\begin{center}\textbf{Abstract}\end{center}\par~~~}%
 {\par}
\def\keywords#1      {\par{\bf Keywords:~}{#1}\par}
\def\pacs#1          {\par{\bf PACS:~}{#1}\par}
\def\subjclass#1     {\par{\bf Subj. class.:~}{#1}\par}

\newcommand\TRASH[1]{}                

\def\Gamma{\varGamma}
\def\Theta{\varTheta}
\def\Lambda{\varLambda}
\def\Xi{\varXi}
\def\Pi{\varPi}
\def\Sigma{\varSigma}
\def\Upsilon{\varUpsilon}
\def\Phi{\varPhi}
\def\Psi{\varPsi}
\def\Omega{\varOmega}
\def\rho{\varrho}

\def\leq{\leqslant}

\def\geq{\geqslant}

\def\SD   {\textrm{\small SD}}
\def\QM   {\textrm{\small QM}}
\def\QC   {\textrm{\small QC}}

\def\NOT  {\textrm{\small NOT}}
\def\CNOT {\textrm{\small CNOT}}
\def\POVM {\textrm{\small POVM}}
\def\DOF  {\textrm{\small DOF}}

\DeclareMathOperator{\Ext}{ext}
\DeclareMathOperator{\erf}{erf}
\newcommand\scprod[2]{\langle{#1},{#2}\rangle}

\def\spH{\mathscr{H}}

\def\grG{\mathscr{G}}
\def\stV{\mathscr{V}}

\def\stE{\mathscr{E}}

\def\qbs{\hat{\sigma}}          
\def\ebs{\hat{e}}               
\def\vbs{\hat{v}}               
\def\vbsl#1{\vbs_{\mathrm{#1}}} 
\def\Xl#1{X^{\mathrm{#1}}}      

\def\tGate{\tau_\mathrm{gate}}
\def\tAvr {\tau_\mathrm{avr}}
\def\tSig {\tau_\mathrm{sig}}

\newcommand\synInh{\unitlength 0.5pt%
 \begin{picture}(18,14)(0,1)%
  \put(0,5){\line(1,0){11}}\put(14,5){\circle*{6}}%
 \end{picture}}
\newcommand\synExc{\unitlength 0.5pt%
 \begin{picture}(18,14)(0,1)%
  \put(0,5){\line(1,0){11}}\thicklines\put(5,5){\vector(-1,0){0}}%
 \end{picture}}
\newcommand\synGat{\unitlength 0.5pt%
 \begin{picture}(18,14)(0,1)%
  \put(0,5){\line(1,0){14}}\put(14,-1){\line(0,1){12}}%
 \end{picture}}
\newcommand\synMod{\unitlength 0.5pt%
 \begin{picture}(18,14)(0,1)%
  \put(0,5){\line(1,0){11}}\put(14,5){\circle{6}}%
 \end{picture}}

\begin{document}
\begin{articlehead}
\begin{raggedright}
\title{Quantum computing in neural networks}
\author{P. Gralewicz}
\address{Department of Theoretical Physics, University of \L\'{o}d\'{z},
  Pomorska 149/153, 90-236, \L\'od\'z, Poland}
\end{raggedright}

\begin{abstract}
According to the  statistical interpretation of quantum theory, quantum
computers form a distinguished class of probabilistic machines (PMs) by encoding
$n$ qubits in $2n$ pbits (random binary variables). This raises the possibility
of a large-scale quantum computing using PMs, especially with neural networks
which have the innate capability for probabilistic information processing.
Restricting ourselves to a particular model, we construct and numerically
examine the performance of neural circuits implementing universal quantum gates.
A discussion on the physiological plausibility of proposed coding scheme is also
provided.
\end{abstract}
\end{articlehead}
\begin{multicols}{2}

\section{Introduction}
  Neural networks are naturally evolved systems for information processing.
Despite decades of experimental and theoretical research, there is no agreement
upon the information encoding employed by these circuits -- the
problem of what exactly is being communicated \textit{via} seemingly chaotic
spike trains is still largely open \cite{RWRB1997}. Advancement in understanding
of this neural language is obstructed by variety of cell types, working
conditions and molecular factors to be taken into account \cite{ChS1992}.
Generally accepted schemes, the \term{rate code} and the \term{phase code}, may
turn out to be only the first two in a sequence of progressively more intricate
codes, where higher order correlations within cellular complexes are utilized.

  Quantum information science, on the other hand, had matured over the last two
decades making significant contributions to both information theory and quantum
mechanics (\QM). The latter, having historical roots in particle physics, is
still often identified with the micro-world. Yet, there is nothing in the
mathematical foundations of {\QM} which could justify that point of view. In
fact, apart from that microscopic realizations, quantum theory has found
many avatars, from mechanical \cite{A1984}, linguistic \cite{AC2003}, purely
geometric \cite{B2004}, to statistical \cite{Bal70,P1995,PeTe98,CFS2002}. In
this article, that last, widely accepted interpretation, is being used to study
the feasibility of a hypothesis that spike trains may actually encode for
quantum states.

  Such hypothesis appears particularly attractive in that the Nature is notorious
in repeating itself at various scales, and if quantum computing (\QC) proves to
be practical, it would be rather surprising if one could not find it implemented
at a higher level. From this point of view neural networks are the obvious
candidates for such implementations. By examining two neural circuits,
designed to perform quantum operations ($1$-qubit rotations, and $2$-qubit
{\CNOT} gate), we demonstrate the feasibility of our hypothesis within the
limits of a simple model. Although quantum registers are realized efficiently
with just two neurons per qubit, the major costs are in the processing of
information carried by the spike trains. The simulations provided are intended
to emphasize the amount of these resources as well as the functionality required
for implementation.

  We begin with a short review of the formalism which allows for the
identification of pairs of spiking neurons with qubits. In Section \ref{sc:ann}
a reduced model of neural network is described, which in Sec. \ref{sc:q-gates}
is further used as a basis for construction of quantum gates. The results of
simulations, in terms of achieved fidelity and coherence, are promising enough
to look toward more realistic implementations. We touch briefly on these issues
in the last section.

\section{Manipulation of quantum states embedded in probabilistic space}
\label{sc:povm}
The operational approach to quantum mechanics, through the formalism of positive
operator-valued measures ({\POVM}s), allows one to
express the states of a quantum system defined in a finite-dimensional
Hilbert space $\spH$, in terms of probability distributions. If the dimension
is $d:=\dim\spH$, then a generic density matrix $\hat{\rho}$ representing the
state has $d^2-1$ degrees of freedom ({\DOF}s). A distribution obtained through
particular {\POVM} has length $d^2$, and -- due to normalization constraint --
the same number of {\DOF}s as the density matrix \cite{PeTe98}. For $n$-qubit
states, this distribution can be associated with joint probability of $2n$
binary random variables\footnote{
  This is in close analogy to complex numbers which extend the reals, and at the
same time are embeddable in a real vector space of doubled dimension equipped
with complex structure.}.

Let $\hat{\rho}$ be a generic density matrix of a $1$-qubit state, which
using summation convention, we write as
\[
  \hat{\rho}
   := \rho^\mu\,\qbs_\mu
    = \begin{pmatrix}
        \thalf + \rho^3 & \rho^1 + \imu \rho^2 \\
        \rho^1 - \imu \rho^2 & \thalf - \rho^3
      \end{pmatrix},
\]
where $\qbs_0=\idmatrix$, $\qbs_{1,2,3}$ are the Pauli matrices,
$\rho^0\equiv \thalf$, and $\rho^1,\rho^2,\rho^3\in[-\thalf,\thalf]$ are the
three real coordinates of a Bloch vector.
Let $\{\hat{A}^{z}\}$ be a normalized $4$-element positive operator-valued
measure 
\begin{equation}
 \label{eq:Anorm1}
  \sum_{z} \hat{A}^{z} = \idmatrix,
  \qquad z=0\Till 3.
\end{equation}
Typically, one associates such a {\POVM} with the Pauli basis, that is
\[
  \hat{A}^{z}
   := A^z\null_\mu \qbs^\mu
    = \tfrac{1}{2}\idmatrix + A^z\null_i \qbs^i,\qquad
  i=1\Till 3.
\]
where $A^z\null_0\equiv \tfrac{1}{2}$, and $\qbs^\mu:=\qbs_\mu^\dagger=\qbs_\mu$,
is the basis dual with respect to the scalar product
$
  \scprod{\qbs^\mu}{\qbs_\nu}
   :=\thalf\tr[\qbs^\mu\qbs_\nu]
    =\delta^\mu\null_\nu
$.
Although not a strict necessity, it is reasonable to assume the same {\POVM} for
all $n$ qubits within a register, and consequently take the entire {\POVM} as a
$n$-fold tensor product
\[
  \hat{A}^{z_1...z_n}
    := \hat{A}^{z_1}\otimes\cdots\otimes\hat{A}^{z_n}.
\]
This leads to the following distribution
\[
 \hspace{-5mm}
 \begin{aligned}
  \null&p^{z_1...z_n}
   := \scprod{\hat{A}^{z_1...z_n}}{\hat{\rho}}
    = 2^{-2n} + A^{z_1}\null_{i_1}\cdots A^{z_n}\null_{i_n}\rho^{i_1...i_n},\\
  \null&\qquad \sum_{z_1...z_n} p^{z_1...z_n} = 1.
 \end{aligned}
\]
Introducing the event basis $\{\ebs_{z}\}$, the transformation can concisely be
written as
\begin{equation}
 \label{eq:rho_p}
  \hat{p} = A \hat{\rho},
\end{equation}
where
\[
 \hspace{-5mm}
 \begin{aligned}
  \hat{p} &= p^{z_1...z_n}\,\ebs_{z_1}\otimes\cdots\otimes\ebs_{z_1},\\
  A &= A^{z_1}\null_{\mu_1}\cdots A^{z_n}\null_{\mu_n}\,
      \ebs_{z_1}\otimes\cdots\otimes\ebs_{z_1}\otimes
      \qbs^{\mu_1}\otimes\cdots\otimes\qbs^{\mu_n}.
 \end{aligned}
\]
Conversely, if $\{\hat{A}^z\}$, are linearly independent, then one can invert
the relation \eqref{eq:rho_p} and take the distribution $\hat{p}$ as an
equivalent representation of the quantum state $\hat{\rho}=A^{-1}\hat{p}$.

A unitary transformation $U\in\grU(2^n)$ of the state is a linear
operator\footnote{
  Note, that the embedding allows to consider a wider range of isometries to
  be implemented, not only the ones corresponding to unitary operations.
  For instance the $1$-qubit antipode (unfortunately also called the
  quantum universal-NOT) can only be approximated in unitary QM
  \cite{BHW1999,MBSS2002}. In probabilistic approach one can realize it exactly.
} $L\in 1\oplus\grSO(2^{2n}-1)$
\[
  \hat{\rho}
   \mapsto U^\dagger\!\hat{\rho}\,U
   = L\hat{\rho},
\]
with elements
\begin{equation}
 \label{eq:L}
 \begin{aligned}
  \null&L^{\mu_1...\mu_n}\null_{\nu_1...\nu_n}\\
  \null&\quad
   = \scprod{\qbs^{\mu_1}\otimes\cdots\otimes\qbs^{\mu_n}}
            {U^\dagger\qbs_{\nu_1}\otimes\cdots\otimes\qbs_{\nu_n}U}.
 \end{aligned}
\end{equation}
After transformation of the basis $A^{-1}:\{\qbs_\mu\}\to\{\ebs_{z}\}$ one has
the same operation acting on probability distribution
\begin{equation}
 \label{eq:Lp}
  \hat{p} \mapsto (A L A^{-1}) \hat{p}.
\end{equation}
There is, however, an important difference between the linear dynamics
of Eq. \eqref{eq:Lp} and Markovian transitions usually considered in association
with stochastic evolution: Denote by $\Omega^{2n}$ the space of joint
probability distributions of $2n$ pbits. Since the operator $A L A^{-1}$ is by
definition invertible, it follows that, in general, it is not a positive one,
hence only a subset of $\Omega^{2n}$ will be mapped back into itself. We denote
this subset -- the (closure of) positive domain of quantum operators -- by
\[
  \overline{\Omega^{2n}_+}:=\{\hat{p}\in\Omega^{2n}\mid A L A^{-1}\hat{p}\in\Omega^{2n}\}.
\]
This is simply the image of all quantum states under the {\POVM} $A$. The
boundary $\Omega^{2n}_0 := \partial\overline{\Omega^{2n}_+}$, which is the image
of the Bloch sphere in $\Omega^{2n}$ contains pure states, while its interior
$\Omega^{2n}_+ := \overline{\Omega^{2n}_+}\setminus\Omega^{2n}_0$
is the subset of mixed states. All remaining distributions
$\Omega^{2n}_- := \Omega^{2n}\setminus \overline{\Omega^{2n}_+}$
are mapped by $A^{-1}$ to the exterior of the Bloch sphere. Therefore, the
{\POVM} partitions the set of possible distributions into three disjoint
subsets:

\par\medskip\begin{tabular}{ll}
 $\Omega^{2n}_0$                     & -- pure quantum states \\
 $\Omega^{2n}_+ = \Int\Omega^{2n}_0$ & -- mixed/decohered states\\
 $\Omega^{2n}_- = \Ext\Omega^{2n}_0$ & -- overcohered states
\end{tabular}\par\medskip

To explain the term \term{overcohered} used above, let us take a closer look
at the limitations imposed by the {\POVM} on distributions in
$\overline{\Omega^{2n}_+}$. Positivity of $A$ implies, that the probabilities
are bound by
\[
  p^{z_1...z_n}\leq 2\avr{p}=2(A^z\null_0\rho^0)^{n}=2^{1-2n}.
\]
Furthermore, if, as we assume, $A$ is non-degenerate, then for any quantum state
only one of the elements $\{p^{z_1...z_n}\}$ can either vanish, or reach the
maximal value $2\avr{p}$. This means, that there is a non-zero lower bound
on the entropy of distributions in $\overline{\Omega^{2n}_+}$, and hence no
distribution with certain outcome can represent a quantum state. Moreover, all
single-pbit marginals are non-vanishing.

  A quantitative characterization of the \term{coherence} can be given by the
radius of the state's Bloch vector. The metric
$g:\Omega^{2n}\times\Omega^{2n}\to\fR$ induced by the
{\POVM} on the distribution space permits to obtain this radius directly for
an arbitrary $\hat{p}\in\Omega^{2n}$. Let
$\hat{p}_1,\hat{p}_2\in\overline{\Omega^{2n}_+}$, and $\hat{\rho}_1$,
$\hat{\rho}_2$ be corresponding quantum states. Then $g$ is given by
\[
  g(\hat{p}_1,\hat{p}_2)
   := \tr[\hat{\rho}_1^\dagger\hat{\rho}_2]
    = (A^{-1}\hat{p}_1)^\dagger\cdot(A^{-1}\hat{p}_2).
\]
Because this is a bilinear map with coefficients independent of
$\hat{p}_1,\hat{p}_2$, one is free to extend its domain onto the entire space
$\Omega^{2n}$. Since $A^0\null_\mu=\thalf$, the radius is
\[
  r^2 := \|\vec{\rho}\,\|^2 = 2^{-n} g(\hat{p},\hat{p})-2^{2n}.
\]
In particular, for a pure state
\[
  r_\mathrm{pure}^2 = 2^{-n}(1-2^{-n}),
\]
and the Bloch radius of any mixed state is always bound by $r<r_\mathrm{pure}$.
The ratio
\begin{equation}
 \label{eq:defR}
  R := \frac{r}{r_\mathrm{pure}},
\end{equation}
can be adopted for a measure of coherence -- ranging from $R=0$ for
maximally decohered state $\hat{\rho}=2^{-n}\idmatrix$, through $R=1$ for any
pure $\hat{\rho}=\hat{\rho}^2$, and beyond $R>1$ for all overcohered ones.

  In order to quantify the performance of circuits considered latter in this
article, we will also employ another, independent measure
by which one can estimate the angular disparity between expected and obtained
states. The \term{fidelity} or normalized overlap between
$\hat{p},\hat{q}\in\Omega^{2n}$ is defined here as
\[
  F := \frac{g(\hat{p},\hat{q})}{\sqrt{g(\hat{p},\hat{p})}\sqrt{g(\hat{q},\hat{q})}}.
\]

  We choose fidelity as a commonly adopted measure, for the purpose of
comparison, notwithstanding direct estimate of the unitary error between
the desired pure state $\hat{p}\in\Omega^{2n}_0$ and obtained distribution
$\hat{q}\in\Omega^{2n}$, which is readily computable:
\[
 \alpha = \arccos\left(\frac{g(\hat{p},\hat{q})-2^{-n}}{r(\hat{q})\sqrt{2^{n}-1}}\right).
\]
The two quantities $\alpha$ and $F$, are nevertheless dependent.

\section{A toy-model neural-network}
\label{sc:ann}
  The information in neural networks is carried by spike trains, which after
appropriate discretization can be transformed to binary strings. The model
network described in this section is a much simplified version of what usually
is considered realistic -- the purpose of such reduction is to retain only the
essential features.
  Consistently with discretization of transmitted signals, the model operates
in explicitly step-wise manner, instead of continuous-time evolution. Likewise,
the delays effected along the inter-neuron paths are also taken to be
integers.

  Let $\grG=(\stV,\stE)$, be a multiply-connected digraph, where
$\stV=\{\vbs_i\}$ is the vertex basis of neurons, (we shall also write
$\stV_{N}$ to explicit the number $N$ of vertices involved) and
$\stE=\stV\otimes\stV\otimes\fN$ is the basis of edges, that is the
possible synaptic connections. The actual couplings between $i^\mathrm{th}$ and
$j^\mathrm{th}$ neuron are set by the weights $W^i\null_{js}$ where $s\in\fN$
enumerates the delays introduced along multiple edges. For each vertex we define
two variables: the binary \term{output state} $X^i\in\{0,1\}$, and the
\term{residual potential} $u^i\in\fR$.

  We adopt the discrete \term{integrate-and-fire} scheme for the dynamics of
this network. In each time step the potential is first updated by accumulating
the incoming signals
\[
  u^{i,t-1}\mapsto u^{it}_\star := u^{i,t-1} + W^i\null_{js}X^{j,t-s}
\]
where the summation runs over connected vertices ($j$) and edge delays
($s\geq 1$).
Subsequent spike generation ($X^{it}=1$) occurs with probability
$P(u^{it}_\star)$, where $P:\fR\to[0,1]$, is a `noisy'
activation function with firing threshold fixed at $u_\mathrm{thr}=\thalf$.
Its actual form used in simulations is given by
\[
  P(u_\star) := \half\left(1+\erf\frac{u_\star - u_\mathrm{thr}}{\sigma}\right),
\]
where $\sigma\geq 0$ is a global control parameter characterizing the noise
standard deviation ({\SD}). In particular, in the limit $\sigma\to 0$ the spikes
are produced deterministically, as $P$ becomes a step-function. The excited
state $u^{it}_\star$ is eventually reduced by release of a spike (refractory
potential), and further quenched with a bound, nonlinear map $S$
\[
  u^{it}_\star \mapsto u^{it} = S(u^{it}_\star- X^{it}).
\]
We assume $S$ to have an attractive fixed point at the origin (the resting
potential),
$
  \forall_u:\;\lim_{t\to\infty} S^t(u)=0
$,
to be linear in its neighborhood
$
  S'(0) = 1
$,
and having finite, but non-zero asymptotes
$
  |S(\pm\infty)|<\infty
$.
The motivation for introduction of this mapping is twofold:
  First, the physiological mechanisms of signal transmission imply existence of
\term{saturations} in both positive and negative direction. The cell can be
depolarized or hyperpolarized through synaptic channels only to certain extent,
and adding more excitatory or inhibitory connections will not have a significant
effect.
  Second, the reason to have $u=0$ for an attractive fixed point, is to
imitate the `leaky' integration scenario, by which in the absence of input
the potential returns back to its resting point.
  In the simulations this function was taken to be a simple, skew-symmetric
mapping
\[
  S(u) := \gamma\tanh\frac{u}{\gamma},\qquad \gamma \geq 0.
\]
Here, the asymptotes are $S(\pm\infty)=\pm\gamma$, therefore we call $\gamma$
the `saturation parameter'. If we assume, that the neuron is left without input and
some residual $u$, so that no spikes are generated, then the potential $u$ will
decay sub-exponentially in time, as
\[
  u \mapsto S^t(u) \underset{t\gg 1}{\to} \frac{\gamma u}{\sqrt{\gamma^2 + \tfrac{2}{3} t u^2}},
\]
where $S^t$ means $t$-fold composition.
In the limit $\gamma\to 0$, the residual potential is reset to zero after each
cycle, and this situation can be associated with time steps longer than the
total refractory time ($\sim 20\,\mathrm{ms}$), within which the cell relaxes to
its resting point. If $\gamma>0$, then the probability of a consecutive spike is
modified by the residual potential: The cell is within the \emph{relative}
refractory period, when the the potassium channels are still open, but the
sodium gates are already reverted to their normal state. This mode corresponds
to time steps of order $\sim 5\,\mathrm{ms}$. Shorter times are generally
unrealistic due to high suppression of spike generation during the
\emph{absolute} refractory period, when the sodium channels are closed.


  The choice of a specific value of $\gamma$ is therefore indirectly related to
the time scale, and consequently to the discretization window of action
potentials. If this window is too short, the discretization becomes
ambiguous and the model breaks down -- this is another reason not to
consider high saturation values.


  The qualitative behavior of the above model is best understood by analyzing
single neuron at the limits of the two control parameters $\sigma$, and $\gamma$.
Assume the cell is fed with a stimulus at a constant frequency
$\nu_\mathrm{in}\in[0,1]$, and consider at first the noiseless regime $\sigma=0$.
If $\gamma=0$, then the only memory of past
input values is stored in delayed connections. The cell fires only if the value
of the convolution $W_{js}X^{j,t-s}$ exceeds the threshold $u_\mathrm{thr}$.
Such neurons acts like a high-pass filter and its firing rate is
$\nu_\mathrm{out}=P(\nu_\mathrm{in}\sum_{s,j} W_{js})$. By increasing the noise
{\SD} $\sigma$, the shape of this filtering function changes along with the
spiking probability $P$, nevertheless it never becomes close to an ideal
multiplier -- the response is always nonlinear.

  If $\gamma\to\infty$ the cell accumulates and `remembers' the residual value
of convolution left over after subtraction of generated spikes. This makes it
into a perfect multiplier with spike rate
$\nu_\mathrm{out}=\nu_\mathrm{in}\sum_{s,j} W_{js}$. Raising $\sigma$ above zero
does not change this average response, but the determinism initially apparent in
the spike patterns is gradually being washed away.

  In between of these two regimes, lies a surprisingly complex area of
fractal-spaced frequency thresholds and output patterns, particularly
conspicuous at $\sigma=0$ and $\nu_\mathrm{in}=1$. Presence of these features,
found in many nonlinear deterministic systems do not critically depend on the
specific shape of the function $S$.

\section{Implementation of universal quantum gates}
\label{sc:q-gates}
  According to the discussion provided in section \ref{sc:povm}, one needs $2n$
random binary variables to implement an $n$-qubit register. In our model
of the neural network, these variables are identified with discretized spikes
registered at $2n$ network sites. The question we set up to address in this
section is, whether there are circuits which can implement state-independent
rotations of the joint probability distributions, that is -- quantum gates.

  The set of gates universal for quantum computation \cite{BBCVMSSSW1995}
includes the whole algebra of $1$-qubit rotations, and an arbitrary $2$-qubit
entangling gate, typically chosen to be the {\CNOT} (controlled-{\NOT}).
Although probabilistic encoding of qubits is efficient ({\ie} linear in $n$),
manipulation of their $2^{2n}$ degrees of freedom ({\DOF}s), by definition
requires exponential amount of resources. From this perspective the construction
of circuits described below should appear at least conceptually straightforward:
The space of binary functions over the vertices $\stV_{2n}$ is
$\stV^{2n}_*=\fZ_{2n}$. We first embed an element $X^t=\{X^{it}\}\in\stV^{2n}_*$
into $\Omega^{2n}$, then apply the gate $G := A L A^{-1}$, and finally project
the result back onto $\stV^{2n}_*$. The entire quantum gate transforming one
set of spike trains $X^t$ to another $Y^t\in\stV^{2n}_*$, is then a composition
\[
  \Pi\circ G\circ\Pi^{-1}:\; X^t \mapsto Y^t,
\]
where $\Pi^{-1}:\stV^{2n}_*\to\Omega^{2n}$, and $\Pi\circ\Pi^{-1}=\id_{\stV^{2n}_*}$.
The main problem in this approach is to construct a reliable projection
$\Pi$, since any information loss during that operation will affect the
quality of entire gate.

  Concrete realization, requires also to decide upon particular {\POVM} being
used. It is possible to choose this transformation in such a way, that some of
the gates will be significantly simplified, for instance acquiring convenient
form of permutations. Our choice is dictated by the optimization of the {\CNOT}
gate, discussed latter in this section. This {\POVM} is given by\footnote{
  From the point of view of state estimation, the optimal POVM is a conformal
 transformation, which maps the Bloch sphere into a sphere inscribed in the
 standard simplex of $\fR^{2^{2n}}$. Thanks to the many symmetries of such
 geometric configuration, some of the rotations are expressible as permutations
 of the simplex' vertices and can be implemented with high efficiency. In the
 case of $A$ given by Eq. \eqref{eq:A}, the permutation $(00,10)(01,11)$
 corresponds to the $1$-qubit NOT gate. With a different POVM 
 one can bring the Hadamard's gate H to a permutation, therefore if an algorithm
 relies on frequent applications of this operation, that could be a preferred
 choice.
}
\begin{equation}
 \label{eq:A}
  A^z\null_\mu := \frac{1}{2}\begin{pmatrix}
      1 &         - \tfrac{1}{\sqrt{3}} &         - \tfrac{1}{\sqrt{3}} &         - \tfrac{1}{\sqrt{3}} \\
      1 &\phantom{-}\tfrac{1}{\sqrt{3}} &\phantom{-}\tfrac{1}{\sqrt{3}} &         - \tfrac{1}{\sqrt{3}} \\
      1 &         - \tfrac{1}{\sqrt{3}} &\phantom{-}\tfrac{1}{\sqrt{3}} &\phantom{-}\tfrac{1}{\sqrt{3}} \\
      1 &\phantom{-}\tfrac{1}{\sqrt{3}} &         - \tfrac{1}{\sqrt{3}} &\phantom{-}\tfrac{1}{\sqrt{3}}
    \end{pmatrix}.
\end{equation}

\subsection{Single-qubit gates}

\begin{figure}
\includegraphics{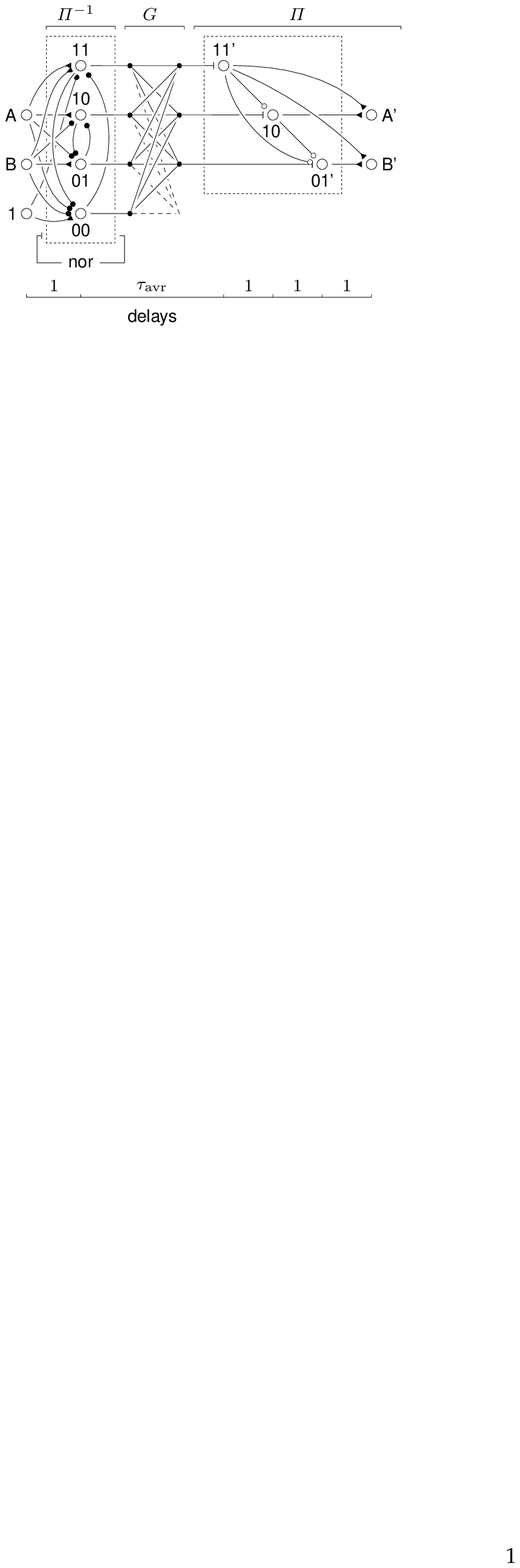}
\figcaption{Schematic for the 1-qubit gate. The hidden nodes are drawn in
  inverted lexicographical order to avoid excessive entanglement of the graph.
  Explicit connections of the normalization (nor) feedback are omitted, to
  improve legibility. Inhibitory connections are marked with (\synInh),
  excitatory with (\synExc), and those capped with (\synGat) depend on the
  applied gate $G$. The double connections (\synMod)
  consist of inhibition followed by delayed and attenuated excitation. The
  transient time is $\tGate = 4 + \tAvr$. Including the normalization feedback,
  and doubled gate connections ($\tAvr=2$), the circuit comprises $10$ nodes,
  and $62$ edges.}
\label{fig:G1D}
\end{figure}

The neural circuit implementing arbitrary $1$-qubit gate is presented in
\figref{fig:G1D}. The projection $\Pi$ which transforms the `sparse' code
$\{\Xl{00'},\Xl{01'},\Xl{10'},\Xl{11'}\}\in\Omega^2$ onto a `dense' one
$\{\Xl{A'},\Xl{B'}\}\in\stV^2_*$, is a linear mapping implemented with
weights
\[
  W_\Pi
    = \begin{pmatrix}
        0 & 0 & 1 & 1 \\
        0 & 1 & 0 & 1
      \end{pmatrix}.
\]
But its inverse, $\Pi^{-1}$ is nonlinear and we realize this function, in a
two-step linear-feedback operation. The first step requires, apart from the
input signals, an additional supply of constant `current' of units from the
vertex $\vbsl{1}$. The effect of such a coupling to unity on a cell is to alter
its firing threshold. The weights of this part, effecting a linear injection
from $\{\Xl{A},\Xl{B},1\}$ to $\{\Xl{00},\Xl{01},\Xl{10},\Xl{11}\}$ are
\[
  W_{\Pi^{-1}} = \begin{pmatrix}
             - 1 &         - 1 &\phantom{-}1 \\
             - 1 &\phantom{-}1 &\phantom{-}0 \\
    \phantom{-}1 &         - 1 &\phantom{-}0 \\
    \phantom{-}1 &\phantom{-}1 &         - 1
    \end{pmatrix}.
\]
While the composition $W_\Pi W_{\Pi^{-1}}=\id_{\stV^2_*}$ as required, the
reciprocal is not an identity and needs a rectifying feedback sent from the
`winning' neuron to its neighbors, in order to bring their residual potentials
back to zero. Because of the one-step delay, this signal has to be adjusted
to match the attenuation already done by the function $S$. Hence the weight
matrix of this rectification step is determined by
\begin{equation}
 \label{eq:Wrec}
 \begin{aligned}
  {[}W_\mathrm{rec}{]}^i\null_j
   &= -S\big([W_{\Pi^{-1}}W_\Pi]^i\null_j - \delta^i\null_j\big)\\
   &= -S(-1)[\idmatrix_{\Omega^2} - W_{\Pi^{-1}}W_\Pi]^i\null_j.
 \end{aligned}
\end{equation}
Note, that for vanishing saturation parameter, this correction also disappears
due to $S\equiv 0$.

\smallskip
  In the absence of noise ($\sigma=0$), the conversion $\Pi^{-1}$ between
dense and sparse coding is completely error-free. As $\sigma$ increases, the
imperfections start to appear in the form of either multiple, or `void' spiking
in the first hidden layer. Although we found the circuit to behave stably in
these conditions, an improvement, in terms of both fidelity and coherence, can
be achieved by adding a second, normalizing feedback (not shown explicitly in
\figref{fig:G1D}).

  Normalizing feedbacks are commonly proposed for explanation of the observed
behavior in cortical neurons \cite{CHM1997,TLDRN1998}. The main difference
between these and our proposal is that while the former are multiplicative, this
one acts additively. Its role is to adjust the residual potentials for the
difference
\[
  1 - \sum_i X^i,\qquad i=00,01,10,11.
\]
Because we do not know which of the four neurons spiked mistakenly, the
normalizing signal is sent evenly to all of them. Its strength is determined by
the average excess of a signal encountered on a double-spike event:
\[
  \tfrac{1}{4}\sum_j [W_\mathrm{rec}]^i\null_j = -\tfrac{1}{4}S(-1).
\]
The deficit, which happens upon lack of a single spike has the same magnitude
but opposite sign. Therefore, the weight matrix of this normalization reads
\[
  W_\mathrm{nor}
    = -S(-1)\begin{pmatrix}
        -\tfrac{1}{4} & -\tfrac{1}{4} & -\tfrac{1}{4} & -\tfrac{1}{4} & 1 \\
        -\tfrac{1}{4} & -\tfrac{1}{4} & -\tfrac{1}{4} & -\tfrac{1}{4} & 1 \\
        -\tfrac{1}{4} & -\tfrac{1}{4} & -\tfrac{1}{4} & -\tfrac{1}{4} & 1 \\
        -\tfrac{1}{4} & -\tfrac{1}{4} & -\tfrac{1}{4} & -\tfrac{1}{4} & 1
      \end{pmatrix},
\]
where the last column refers to the unit vertex $\vbsl{1}$.

\smallskip
  Making the embedding $\Pi^{-1}$ robust is crucial for achieving correct
projection $\Pi$. Multiple, or void events linearly projected \textit{via}
$W_\Pi$ typically lead to a significant loss of coherence (although with less
impact on fidelity). In order to compensate for these non-exclusive events, the
spiking neuron sends a composite inhibitory signal down the hierarchy.
This is implemented with double connections: the first transmits
inhibitory signal at certain level $\eta$, and is followed by a delayed,
excitatory one aimed at bringing the residual potential of target neuron back
to its prior value. The full accomplishment of this goal is impossible with
nonlinear function $S$ -- the value can only be fully restored in the linear
limit $\gamma\to\infty$. Given the inhibitory coupling $\eta$, our best estimate
of the following excitation strength is $\eta'=\avr{u} - S(\avr{u}-\eta)$, where
$\avr{u}$ is the average residual potential. Because $\avr{u}\approx 0$, we set
$\eta'=-S(-\eta)$. The optimal value of $\eta$ was found numerically, by
minimizing the variation of fidelity over a range of gates acting on test
states (see Results).

\smallskip
  Application of a gate $G$ requires no additional node of the network,
only manipulation of the weights between the embedding and projecting parts.
In simplest case these are directly set to
\[
  W_G = A L A^{-1} = G.
\]
We have found however, that within some limits, the mechanism of \term{synaptic
averaging} may provide improvement of the performance. In real networks, a
single synapse contributes only a tiny fraction of the total input signal
\cite{ChS1992}. Multiple connections of similar lengths lead to signal
accumulation, different delays -- to temporal averaging. In our toy-model, the
first case is replaced by single, but strong connections, while for
implementation of the latter we directly use several edges having different
lengths with proportionally attenuated couplings. In the case of a single-qubit
gate, of the several configurations tested, the best results were obtained with
just two-step average $\tAvr=2$, hence, the connections were fixed at
$W_{G,1}=W_{G,2}=\thalf G$.

\begin{figure}
\includegraphics{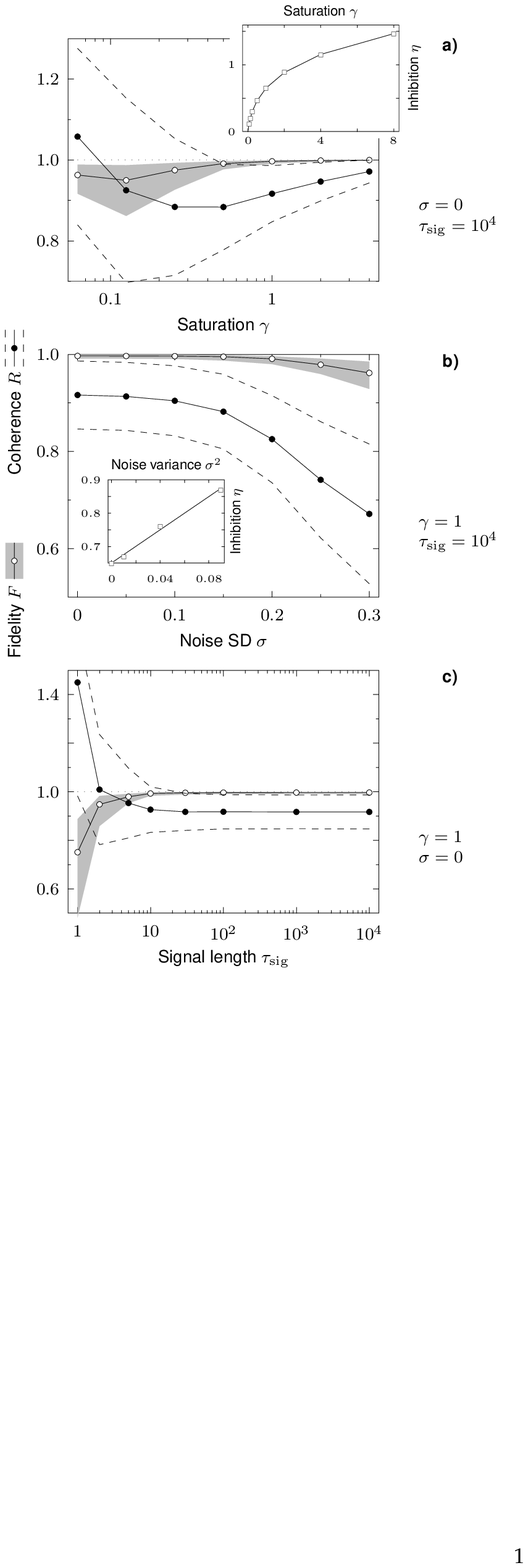}
\figcaption{Performance of the 1-qubit phase gate in function of \textbf{a)}:
  saturation level of the residual potential, \textbf{b)}: noise SD of the
  activation function $P$, \textbf{c)}: length of the input signal (note the
  scale difference between graphs). Synaptic averaging was fixed at $\tAvr=2$.
  Each point is a mean over $36$ rotation angles within the whole interval
  $[0,2\pi)$, and $14$ pure states on which the phase gate was tested. With
  statistics of $10^4$ steps per setting, the associated uncertainties are
  negligible -- shaded regions ($F$) and broken lines ($R$) represent the
  standard deviations across the states and rotation angles.
  \textbf{Insets}: The inhibition levels used during simulations, optimized for
  minimization of the fidelity variance.}
\label{fig:FC-1D}
\end{figure}

\paragraph{Results.}
  In order to reduce statistical uncertainties, all gates were tested on a fixed
set of $14$ pure states approximately evenly distributed on the Bloch sphere:
\[
 \hspace{-1mm}
 \begin{aligned}
  \Big\{&\ket{0}, \ket{1},
        \tfrac{1}{\sqrt{2}}\big(\ket{0}+\e{\imu k\pi/2}\ket{1}\big),\\
  \null&\frac{\sqrt{\sqrt{3}\pm 1}\ket{0}+\e{\imu(2k+1)\pi/4}\sqrt{\sqrt{3}\mp 1}\ket{1}}
             {\sqrt{2\sqrt{3}}}\Big\},
 \end{aligned}
 \quad
 k \in \fZ_4.
\]
While the input nodes were fed with spike trains $\{\Xl{A},\Xl{B}\}$ of joint
probability distributions $\hat{p}_\mathrm{in}\in\Omega^2_0$ corresponding to
the above states, the output $\{\Xl{A'},\Xl{B'}\}$ was tested for its coherence
$R$, and fidelity with the desired distribution $G\hat{p}_\mathrm{in}$.
  The initial test runs were made for several gates including Hadamard's, {\NOT},
the antipode (non-unitary), and two rotations:
$U_\theta = \exp(\imu\theta\hat{\sigma}_2/2)$, and
$U_\phi   = \exp(\imu\phi\hat{\sigma}_3/2)$. For the representative, the phase
gate $U_\phi$ was selected -- the effects of other operations were
quantitatively similar, or better. Its representation $G_\phi=A L_\phi A^{-1}$,
acting in $\Omega^2$ reads
\[
 \hspace{-7mm}
  G_\phi
    = \half\begin{pmatrix}
        1 + \cos\phi & 1 - \cos\phi &-\sin\phi     & \sin\phi     \\
        1 - \cos\phi & 1 + \cos\phi & \sin\phi     &-\sin\phi     \\
        \sin\phi     &-\sin\phi     & 1 + \cos\phi & 1 - \cos\phi \\
       -\sin\phi     & \sin\phi     & 1 - \cos\phi & 1 + \cos\phi
      \end{pmatrix}.
\]
The results presented in \figref{fig:FC-1D} are averages over $36$
rotation angles evenly spaced across the entire interval $[0,2\pi)$. The best
performance was observed for $\phi=0$ (identity) and $\phi=\pi$, while the worst
cases were encountered around $\phi\approx \pm \pi/2$ (but not exactly at these
angles). For each setting $(\sigma,\gamma)$, the inhibition level $\eta$ was
adjusted to minimize the variance of fidelity across the test states and
rotation angles ({\cf} \figref{fig:FC-1D}-insets). Note that while this
optimization was mainly coincident with maximization of the fidelity itself,
the trend in coherence was typically opposite. Had we chosen to optimize for
purity of states ($R\to 1$), the figures would look different.

  The prominent feature of \figref{fig:FC-1D}a, is the overcoherence of output
states in the limit $\gamma\to 0$. This means these distributions are too sharp
to represent quantum states, and any subsequent application of another gate
would certainly lead to a loss of accuracy. Interestingly, the
average fidelity remains at relatively high level. This suggests a possibility
of correcting the distributions by rescaling about the average. On the other
hand, the fidelity {\SD} is significant for small saturations, and becomes
comparable with statistical uncertainties only above $\gamma\gtrsim 1$.

  The conclusion drawn from \figref{fig:FC-1D}b is clear: the circuit considered
here is designed to work in deterministic regime $\sigma\to 0$. This makes an
interesting contrast between stochastic nature of quantum states and the
determinism of gates acting on them. As we are going to show, this dichotomy is
not limited to the $1$-qubit gate, but persists also in the case of entangling
operation {\CNOT}.

  Finally, we have sought for an estimate of the time needed to complete the
quantum rotations with this gate. Apart from the spatial resources, measured in
terms of cells and connections being used, time is an important factor
contributing to the overall cost of the realization. To assess this property, we
have run the circuit while varying the \term{signal length} $\tSig$: After an
initial transient of $\tGate = 4 + \tAvr$, the network was ran for $\tSig\geq 1$
successive steps, after which the cells were re-set to their initial state
($u^i=0$, $X^i=0$), ensuring that all memory traces stored in residual
potentials were erased. This procedure was repeated until satisfactory
statistics ($N\tSig\approx 10^4$) was gathered.

  The results provided in \figref{fig:FC-1D}c evidently show that the real
temporal cost is not only the delay $\tGate$, but a significant number of
further steps are needed to `tune' this gate to a signal. After approximately
$\tSig\approx 30$ events the output quality no longer improves, and
consequently one can identify $\tSig$ with the statistics needed for maximal
efficiency. Since the latter is a function of saturation $\gamma$ and noise
$\sigma$, one expects $\tSig$ to raise monotonically with $\gamma$ and decrease
as $\sigma$ increases. In particular, the ideal case $\gamma\to\infty$,
$\sigma\to 0$ would also require infinite statistics to achieve the best
performance. One therefore finds yet another reason for the low saturation
values: The finiteness of signals encoded in spike trains, limits the
attainable efficiency of transformations, and high saturation values cannot
provide improvement beyond these limitations.

\subsection{The CNOT gate}
\begin{figure}
\hspace{-3mm}\includegraphics{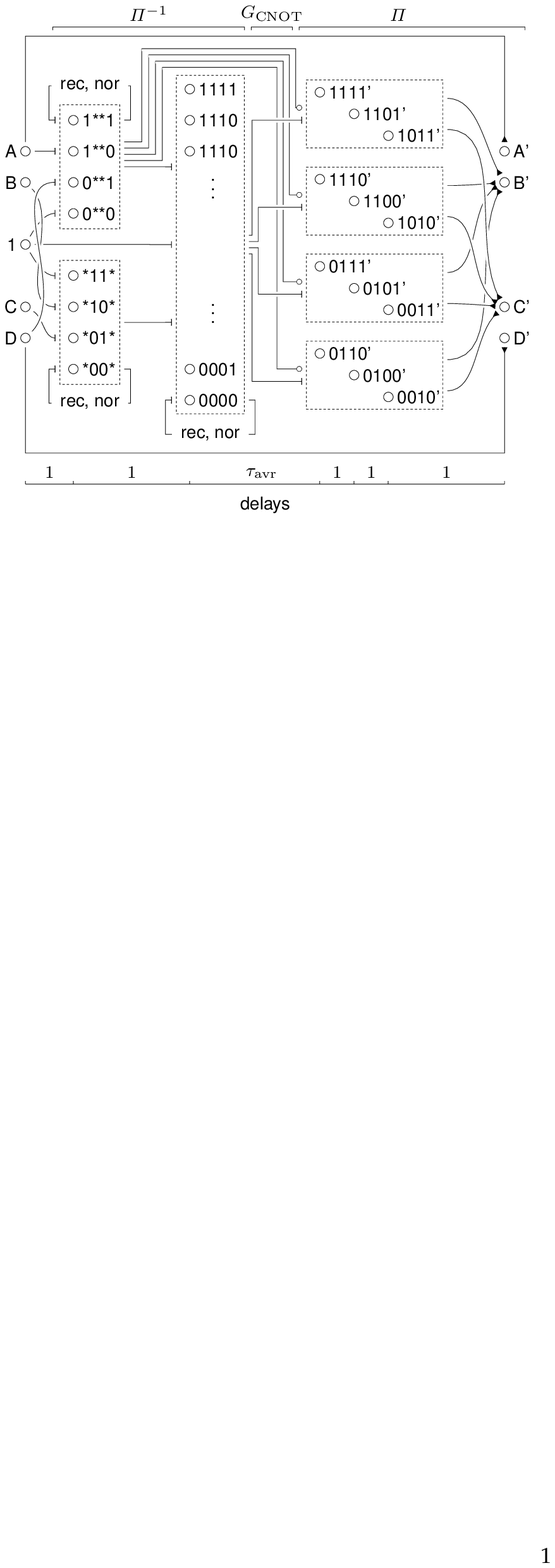}
\figcaption{Schematic of the CNOT gate. The circuit has $38$ nodes, and $1309$
  edges if the synaptic averaging is set to $\tAvr=4$. The transient time is 
  $\tGate = 5 + \tAvr$.}
\label{fig:CNOT}
\end{figure}

Unlike the single-qubit gates which can, by means of a special choice of the
{\POVM}, be transformed to a permutation, the {\CNOT} operation does not admit
such representation\footnote{
  Would it be possible, then either the gate could have no entangling
  capability, or the marginal probabilities of the two qubits be not conserved.
}.
With $A$ given by \eqref{eq:A}, its operator $G_\mathrm{CNOT}=A L_\mathrm{CNOT} A^{-1}$ has the following
structure
\[
 \hspace{-9mm}
  G_\mathrm{CNOT}
    = \begin{pmatrix}
        H_1 - J_1            & H_1^\mathrm{T} + J_2 & H_2                  & H_2^\mathrm{T}       \\
        H_1^\mathrm{T} + J_2 & H_1 - J_1            & H_2^\mathrm{T}       & H_2                  \\
       -H_2                  &-H_2^\mathrm{T}       & H_1 - J_2            & H_1^\mathrm{T} + J_1 \\
       -H_2^\mathrm{T}       &-H_2                  & H_1^\mathrm{T} + J_1 & H_1 - J_2
      \end{pmatrix},
\]
where
\begin{align*}
  H_1 &= \frac{1}{4}\begin{pmatrix}
          1 & \tfrac{1}{\sqrt{3}} & 1 & \tfrac{1}{\sqrt{3}}\\
         -\tfrac{1}{\sqrt{3}} & 1 &-\tfrac{1}{\sqrt{3}} & 1\\
          1 & \tfrac{1}{\sqrt{3}} & 1 & \tfrac{1}{\sqrt{3}}\\
         -\tfrac{1}{\sqrt{3}} & 1 &-\tfrac{1}{\sqrt{3}} & 1
        \end{pmatrix},\\
  H_2 &= \frac{1}{4}\begin{pmatrix}
          \tfrac{1}{\sqrt{3}} &-1 &-\tfrac{1}{\sqrt{3}} & 1\\
          1 & \tfrac{1}{\sqrt{3}} &-1 &-\tfrac{1}{\sqrt{3}}\\
         -\tfrac{1}{\sqrt{3}} & 1 & \tfrac{1}{\sqrt{3}} &-1\\
         -1 &-\tfrac{1}{\sqrt{3}} & 1 & \tfrac{1}{\sqrt{3}}
        \end{pmatrix},\\
  J_1 &= \frac{1}{\sqrt{3}}\begin{pmatrix}
          1 & 0 & 0 & 0 \\
          0 & 0 & 0 &-1 \\
          0 & 0 & 1 & 0 \\
          0 &-1 & 0 & 0
        \end{pmatrix},\\
  J_2 &= \frac{1}{\sqrt{3}}\begin{pmatrix}
          0 & 0 & 1 & 0 \\
          0 &-1 & 0 & 0 \\
          1 & 0 & 0 & 0 \\
          0 & 0 & 0 &-1
        \end{pmatrix}.
\end{align*}
As already mentioned, the {\POVM} \eqref{eq:A} has been
chosen to optimize the {\CNOT} gate. Indeed, the linear projection
$\Pi:\Omega^4\to\stV^4_*$,
\[
  W_\Pi =
    \begin{pmatrix}
      0 & 0 & 0 & 0 & 0 & \ldots & 1 & 1 \\
      0 & 0 & 0 & 0 & 1 & \ldots & 1 & 1 \\
      0 & 0 & 1 & 1 & 0 & \ldots & 1 & 1 \\
      0 & 1 & 0 & 1 & 0 & \ldots & 0 & 1
    \end{pmatrix},
\]
when combined with $G_\mathrm{CNOT}$, 
shows the two pbits $\Xl{A}$ and $\Xl{D}$ are invariant under $G_\mathrm{CNOT}$.
These can be directly copied to the output $\Xl{A'}$, $\Xl{D'}$, as shown in
\figref{fig:CNOT}. For that same reason, while implementing the embedding part
$\Pi^{-1}$, we pair $\{\Xl{A},\Xl{D}\},\{\Xl{B},\Xl{C}\}$, rather than using the
natural order $\{\Xl{A},\Xl{B}\},\{\Xl{C},\Xl{D}\}$. The section $\Pi^{-1}$ is
a two-stage procedure: First, with the same construction as in $1$-qubit case
we separately embed the two marginals  $\{\Xl{A},\Xl{D}\}$ and
$\{\Xl{B},\Xl{C}\}$:
\[
  \Pi^{-1}_\mathrm{I}:\stV^4_*\to\Omega^2\times\Omega^2.
\]
Next, we combine these into a single map
\[
  \Pi^{-1}_\mathrm{II}:\Omega^2\times\Omega^2\to\Omega^4.
\]
Since this is done with a linear mapping, there is again a rectifying feedback
$W_\mathrm{rec,II}$ obtainable from Eq. \eqref{eq:Wrec} applied to
$\Pi_\mathrm{II}$ on $\Omega^4$. In contrast to $\Pi^{-1}_\mathrm{I}$, the
second-stage embedding $\Pi^{-1}_\mathrm{II}$ turns out to be unstable against
noise, and the normalization feedback is now a necessity. Because of
\[
  \forall_i\; 2^{-4}\sum_j [W_\mathrm{rec,II}]^i\null_j =-\tfrac{9}{16} S(-1),
\]
the normalization weights are set to
\[
  W_\mathrm{nor,II} = -9S(-1)(-D_4\oplus 1),
\]
where $[D_4]^i\null_j\equiv 2^{-4}$ is the diffusion operator, and `$1$' refers
to the unit vertex $\vbsl{1}$.

  Thanks to the invariance of two pbits $\Xl{A'}=\Xl{A}$, $\Xl{D'}=\Xl{D}$, the
hierarchical projection $\Pi$ have been significantly simplified (in comparison
to what is needed for general $2$-qubit gate). The four partial projections
from $G_\mathrm{CNOT}$, shown on the right hand side of \figref{fig:CNOT}, are
modulated directly by the marginal $\Pi^{-1}_\mathrm{I}(\Xl{A},\Xl{D})$. The
mechanism of this modulation is the same as explained before (inhibition
followed by attenuated excitation) and the same parameter $\eta$ is set common
on those connections.

  Finally, the gate edges were multiplied, in order to use the synaptic
averaging mechanism. We found no dramatic improvement while varying the
averaging length $\tAvr$, at $\gamma\gtrsim 1$, nevertheless the
performance was significantly better for small values of the saturation
parameter. At $\gamma=1$ the optimal length was $\tAvr=4$.

\paragraph{Results.}
  The performance was assessed upon a testing set of $28$ pure states, which
included both separable and entangled ones:
\[
 \hspace{-9mm}
 \begin{aligned}
  \big\{
   &\ket{00},\ket{01},\ket{10},\ket{11},\\
   &\tfrac{1}{\sqrt{2}}\big(\ket{00}+\e{\imu k\pi/2}\ket{01}\big),\,&
   &\tfrac{1}{\sqrt{2}}\big(\ket{00}+\e{\imu k\pi/2}\ket{10}\big),\\
   &\tfrac{1}{\sqrt{2}}\big(\ket{00}+\e{\imu k\pi/2}\ket{11}\big),\,&
   &\tfrac{1}{\sqrt{2}}\big(\ket{01}+\e{\imu k\pi/2}\ket{10}\big),\\
   &\tfrac{1}{\sqrt{2}}\big(\ket{01}+\e{\imu k\pi/2}\ket{11}\big),\,&
   &\tfrac{1}{\sqrt{2}}\big(\ket{10}+\e{\imu k\pi/2}\ket{11}\big)
  \big\},
 \end{aligned}
 \quad
 k\in\fZ_4.
\]
Interestingly, although some of these states are `preferred' in terms of
achieved fidelity $F$, there was no correlation between this measure and the
entanglement property. This observation should not be surprising, because the
mapping of $2$-qubit states into joint probability distributions does not make
entangled states distinct. It follows that even imperfect gate implementation
should not distinguish these states from separable ones.

\begin{figure}
\includegraphics{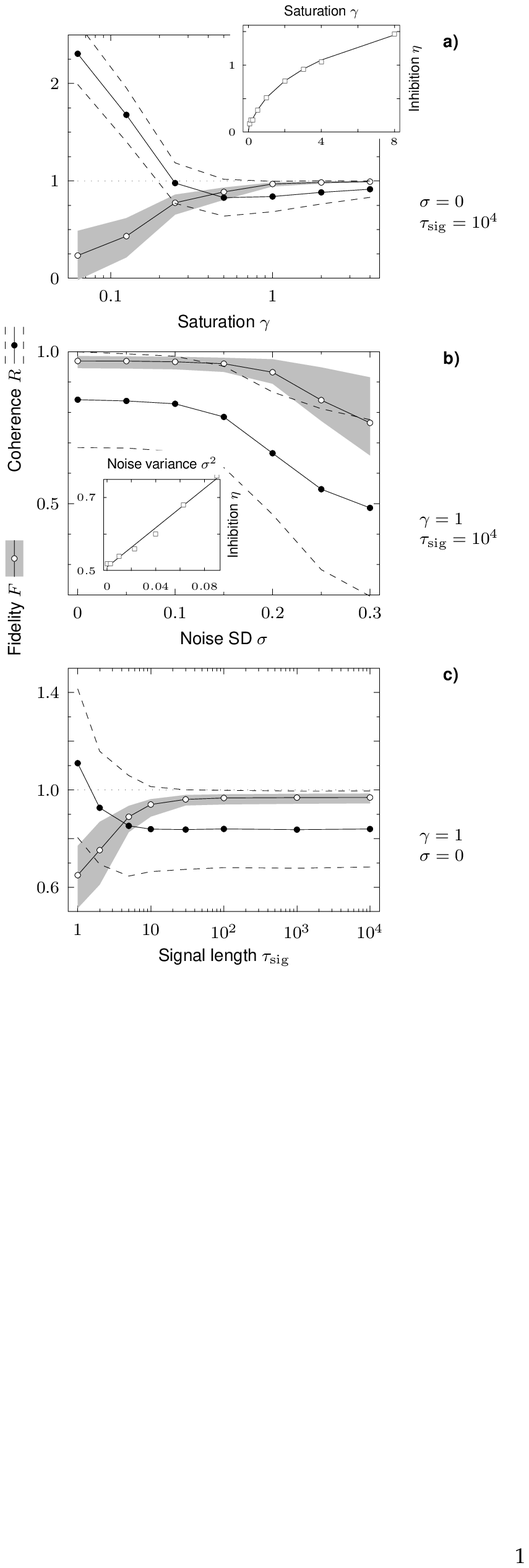}
\figcaption{Performance of the CNOT gate for synaptic averaging fixed at
  $\tAvr=4$. Note the differences in ranges, while comparing with
  Fig.~2. The statistics for each setting is $10^4$ steps, and each point is an
  average over $28$ test states.}
\label{fig:FC-CNOT}
\end{figure}

The results of simulations are presented in \figref{fig:FC-CNOT}. Like before,
for each setting of the control parameters $(\gamma,\sigma)$, the inhibition
level $\eta$ was adjusted to minimize the variance of fidelity across the
test states.

Qualitatively, the figures \ref{fig:FC-CNOT} are largely similar to what had
been obtained for $1$-qubit gates (\figref{fig:FC-1D}), the major difference
is in the range of achieved fidelities and output coherences. Standard deviations
of coherence $R$ increased evenly by approximately a factor of $2$, while the
fidelity {\SD} multiplied by about $4-5$. The most dramatic changes are
observed in \figref{fig:FC-CNOT}a: Whereas at low saturation values
($\gamma<1$) the $1$-qubit gate worked relatively well, in the case of
{\CNOT} a huge overcoherence takes place along with significant fidelity
loss.

  For a reasonable performance at $\sigma=0$ and $\tSig \gtrsim 30$ one
needs $\gamma \gtrsim 1$. In this regime one finds $F\gtrsim 0.97\,(-0.03,+0.02)$,
corresponding to the unitary error $\alpha \lesssim 14^\circ\,(-5,+4)$; with
noise at $\sigma=0.3$ and $\gamma=1$ the fidelity drops down to
$F=0.77\,(-0.11,+0.15)$, or $\alpha=42^\circ\,(-22,+12)$, what is hardly
acceptable for a large-scale quantum computation. While comparing these
values with the best to-date experimental achievements ($F \sim 0.7-0.8$ with
trapped ions \cite{SKHRGLDBREB2003}, $F \sim 0.6-0.8$ with Josephson junctions
\cite{YPANT2003}, $F \sim 0.85$ in optical setup \cite{BPWR2003}) one has to
take into account the many simplifications of our toy-model. More realistic
simulations, or ultimately -- realizations, may not necessarily prove as good as
this one, and would probably require additional resources to implement some of
the error correcting schemes, nevertheless the principle of quantum computing
with neural networks has been demonstrated.

\section{Discussion}
\label{sc:done}

  We have studied the potential of an artificial neural network to operate on
correlated spike trains assuming the latter to encode quantum states. The model
neurons are reduced here down to the essential ingredients of computational
capability. Few comments concerning the simplifications made are in order at
this point:

  First, we have completely neglected the synaptic noise, by assuming the
signals to be relayed undisturbed between cells. The justification is that
here the few edges of each node represent averages over $10^3-10^4$ real
synaptic connections therefore the impact of faulty transmission through a
single synapse is greatly reduced. But inclusion of this likely source of errors
may still be a significant factor reducing the overall performance.

  Second, the time duration of processed signals are assumed to be much shorter
than the synaptic plasticity scale. Adaptation is an inherent element of
information processing in the brain, but it conflicts with the objective of
reliable signal transformations in that there is a trade-off between computing
efficiency and adaptive capability. The resolution is provided by separation in
time scales between the two processes -- transformations act over short signals,
typically in response to rapidly varying external stimuli. This is consistent
with the optimal signal length which was found here for both $1$- and $2$-qubit
gates to be of order $\sim 30$ steps. Assuming the time step is set to
$\sim 5\,\mathrm{ms}$ leads to a realistic signal duration of
$\sim 150\,\mathrm{ms}$.

  Third, the detrimental effect of cellular noise on the performance of quantum
gates clearly shows the deterministic regime to be preferable at least for the
coding scheme considered here. On one hand, a sharp firing threshold needed for
the neurons to act as `counters' which discretize linearly accumulated input
signals, corroborates with the theoretical analysis of optimality in terms of
information encoding \cite{BRP2003}.
  But on the other, the noise itself which blurs this
threshold has been shown to be a viable resource acting through the mechanisms
of stochastic resonance \cite{GHJM1998}. This suggests to consider alternative
quantum coding schemes, which would make use of the inherent uncertainty in spike
generation, provided the relevant conditions are stable enough ({\eg}, noise
variance at a constant, moderate level).

  It is worthwhile to note at this point, that the quantum states are not
absolute entities, and the same set of spike trains may be `quantized' in many
different ways depending on the assumed definition of a state. Accordingly, the
quantum transformations as well as their implementations will differ. We have
discussed here only two coding schemes (referred to as the `dense' and `sparse'
spatial code), but it appears plausible, that the real networks may actually
alternate (or combine) many different encodings, depending on the nature of
the input signal and the functional properties of the circuit. An evident
possibility is the \emph{sparse temporal code} based on probability waves,
particularly attractive for at least two reasons: First, the brain waves
provide the frequency basis necessary for phase discrimination, and there is an
experimental indication for independence between rate and phase variables
\cite{HBK2003}. The question is not whether the spiking probability oscillation
does have a role, but rather what is the relevant number of modes involved
in computation (if more than two then one should consider qudits instead of just
qubits).
Second, while the `dense' code requires two random binary variables per qubit,
by trading spatial for temporal resources, probability waves allow to encode
one-qubit per neuron.
  The drawback is that the mechanisms of short term synaptic plasticity
\cite{MLFS1997,BP1998,FD2002} makes the neural circuits operating on this form
of a code susceptible to unwanted modifications \cite{FLBR2001}. From this
perspective, the use of sparse spatial coding \cite{F1987,BNMLL1990,OF1996,ASGA1996,BS1997,KBTGA2003},
appears to be advantageous, since such spike trains have by definition no
temporal correlations, and hence the circuits operating in this fashion are
expected to be more stable.

\smallskip
  In summary, we have demonstrated the principle of employing quantum
coding in artificial neural networks, by providing examples of circuits which
realize quantum gates. There is a room for improvement and further investigation
with more realism put into the model, alternative circuits, and algorithm
implementations. Exploring the possible ways in which neural networks can handle
quantum codes, can certainly benefit both the quantum mechanics and neuroscience.
On one hand, applications of {\QM} to neural systems broaden the range of
possibilities to be considered when seeking to understand the language of
spikes, on the other -- macroscopic realizations can provide clues about the
microscopic phenomena upon which {\QM} originated.

\def\bbaut#1#2{{#1}~{#2},}
\def\bbttl#1{\textsl{``#1''},}
\def\bbpub#1{\textit{#1}}

\clearpage
\end{multicols}
\end{document}